\documentclass[preprint,12pt]{elsarticle}

\usepackage{graphicx}
\usepackage{epsfig}
\usepackage{array}
\usepackage{amscd,amsmath,amssymb}

\def\kcnn{K^{\pm} \rightarrow \pi^\pm \pi^0 \pi^0}
\def\kccc{K^{\pm} \rightarrow \pi^\pm \pi^+ \pi^-}
\def\p0p0{\pi^0 \pi^0}
\def\pp{\pi^+ \pi^-}

\def\mm2{M_{00}^2}

\def \geve     {\mbox{$\mathrm{GeV}$}}
\def \gevp     {\mbox{$\mathrm{GeV}/c$}}
\def \gevm    {\mbox{$\mathrm{GeV}/c^2$}}
\def \gevmsq {\mbox{$(\mathrm{GeV}/c^2)^2$}}

\begin{document}

\begin{frontmatter}

\title{Empirical parameterization of the $\kcnn$ decay \\ Dalitz plot \\
\begin{center}
\small
The NA48/2 Collaboration
\end{center}
}

\author[a0]{J.R.~Batley}

\author[a0]{A.J.~Culling}

\author[a0]{G.~Kalmus}

\author[a0,f1]{C.~Lazzeroni}

\author[a0]{D.J.~Munday}

\author[a0,f1]{M.W.~Slater}

\author[a0]{S.A.~Wotton }

\address[a0]{Cavendish Laboratory, University of Cambridge, Cambridge, CB3 0HE, UK$^1$} 

\author[a1,a17,a18]{R.~Arcidiacono}

\author[a1]{G.~Bocquet}

\author[a1,f4U,f4Z]{N.~Cabibbo}

\author[a1]{A.~Ceccucci}

\author[a1,f5]{D.~Cundy}

\author[a1]{V.~Falaleev}

\author[a1]{M.~Fidecaro}

\author[a1]{L.~Gatignon}

\author[a1]{A.~Gonidec}

\author[a1]{W.~Kubischta}

\author[a1,a5D,a5Z]{A.~Norton}

\author[a1]{A.~Maier}

\author[a1]{M.~Patel}

\author[a1]{A.~Peters}

\address[a1]{CERN, CH-1211 Gen\`eve 23, Switzerland} 

% ...

\address[a19]{\"Osterreichische Akademie der Wissenschaften, Institut f\"ur Hochenergiephysik, A-10560 Wien, Austria$^4$} 

\address[f1]{University of Birmingham, Edgbaston, Birmingham, B15 2TT, UK}

\address[f4U]{Universit\`a di Roma ``La Sapienza'', I-00185 Roma, Italy}

\address[f4Z]{Sezione dell'INFN di Roma, I-00185 Roma, Italy}

\address[f5]{Istituto di Cosmogeofisica del CNR di Torino, I-10133 Torino, Italy}

\address[f9]{Faculty of Physics, University of Sofia ``St. Kl. Ohridski'', 5 J. Bourchier Blvd., 1164 Sofia, Bulgaria}


\address[f12]{Centre de Physique des Particules de Marseille, IN2P3-CNRS, Universit\'e de la M\'editerran\'ee, Marseille, France}

\address[f13]{Department of Physics and Astronomy, George Mason University, Fairfax, VA 22030, USA}

\address[f14]{Dipartimento di Fisica, Universit\`a di Modena e Reggio Emilia, I-41100 Modena, Italy}

\address[f15]{Istituto di Fisica, Universit\`a di Urbino, I-61029 Urbino, Italy}

\address[f16]{Physikalisches Institut, Universit\"at Bonn, D-53115 Bonn, Germany}

\address[f18]{SLAC, Stanford University, Menlo Park, CA 94025, USA}

\address[f19]{Royal Holloway, University of London, Egham Hill, Egham, TW20 0EX, UK}

\address[f20]{UCLA, Los Angeles, CA 90024, USA}

\address[f21]{Laboratori Nazionali di Frascati, I-00044 Frascati (Rome), Italy}

\address[f22]{Institut de F\'isica d'Altes Energies, UAB, E-08193 Bellaterra (Barcelona), Spain}

\address[f24]{University of Bern, Institute for Theoretical Physics, Sidlerstrasse 5, CH-3012 Bern, Switzerland}

\address[f25]{Centro de Investigaciones Energeticas Medioambientales y Tecnologicas, E-28040 Madrid, Spain}

\cortext[cor1]{Corresponding author}

\begin{abstract}
As first observed by the NA48/2 experiment at the CERN SPS, the $\p0p0$ invariant 
mass ($M_{00}$) distribution from $\kcnn$ decay shows a cusp-like anomaly at 
$M_{00}=2m_+$, where $m_+$ is the charged pion mass. 
An analysis to extract the $\pi\pi$ scattering lengths in the isospin
$I=0$ and $I=2$ states, $a_0$ and $a_2$, respectively, has been recently reported. 
In the present work the Dalitz plot of this decay is fitted to a new empirical 
parameterization suitable for practical purposes, such as Monte Carlo simulations
of $\kcnn$ decays.
\end{abstract}


\end{frontmatter}

\footnotetext[1]{Funded by the UK Particle Physics and Astronomy
Research Council}

\footnotetext[2]{Funded by the German Federal Minister for
Education and research under contract 05HK1UM1/1}

\footnotetext[3]{Funded by the German Federal Minister for Research
and Technology (BMBF) under contract 056SI74}

\footnotetext[4]{Funded by the Austrian Ministry for Traffic and
Research under the contract GZ 616.360/2-IV GZ 616.363/2-VIII, and
by the Fonds f\"ur Wissenschaft und Forschung FWF Nr.~P08929-PHY}


\section{Introduction}
\label{intro}

Since 1960 the square of the matrix element absolute value $|M|$ which
describes the $\kcnn$ Dalitz plot distribution has been parameterized by a 
series expansion such as that introduced by Weinberg \cite{Weinberg:1960zza}:
\begin{equation} 
\frac{d |M|^2 }{dU dV} \propto 1 + GU + HU^2 +KV^2 + ... ,
\label{old}
\end{equation}
where $U = (s_3 - s_0)/m_{\pi^+}^2$, $V = (s_2 - s_1)/m_{\pi^+}^2$ and
$$s_i = (P_K - P_i)^2, i=1,2,3; \ \ \ s_0=(m_{K^+}^2 + 2 m_{\pi^0}^2 + m_{\pi^+}^2)/3.$$
Here $P_i$ are the $i^{th}$ pion four-momenta and $i=3$ is assigned to the charged pion.
The latest measurements of the $G$, $H$ and $K$ parameters using Eq. (\ref{old}) are published in 
\cite{Ajinenko:2002mg, Akopdzhanov:2005nb}, and the corresponding PDG average values 
\cite{Amsler:2008zz} are $G=0.626 \pm 0.007, H = 0.052 \pm 0.008, K = 0.0054 \pm 0.0035$.

However, in 2005 the NA48/2 experiment at the CERN SPS first observed a cusp-like 
anomaly in the  $\p0p0$ invariant mass $(M_{00})$ distribution of this decay
in the region around $M_{00}= 2m_+$, where $m_+$ is the charged pion mass  
\cite{Batley:2005ax}.  This anomaly had been predicted in 1961
\cite{Budini:1961} as an effect due mainly to the destructive interference between the 
direct amplitude of $\kcnn$ decay and the final state charge exchange scattering process 
$\pp \rightarrow \p0p0$ in $\kccc$ decay (see also \cite{Cabibbo:2004gq}).
 
Best fits using two theoretical formulations of rescattering effects \cite{Cabibbo:2005ez} and
\cite{Colangelo:2006va, Bissegger:2008ff} have provided a precise determination
of  $a_0 - a_2$, the difference between the S-wave 
$\pi\pi$ scattering lengths in the isospin $I=0$ and $I=2$ states, and an 
independent, though less precise, determination of $a_2$ \cite{Batley:2009cusp}. Such an analysis
leads to a successful fit of the Dalitz plot using a rather long expression which contains
physically meaningful constants (that could be measured better in future) 
and is affected by theoretical uncertainties. It is not practical to
implement these formulae if one just needs to describe 
the Dalitz plot shape, say, in a Monte Carlo generator. On the other hand,
it is known now \cite{Batley:2005ax} that the Dalitz plot region near $s_3 = (2m_+)^2$ cannot be described 
by Eq. (\ref{old}), so a model-independent, empirical description of
$\kcnn$ decay is certainly useful to replace the old parameterization.

The main purpose of the  NA48/2 experiment at the CERN SPS was to search for
direct CP violation in $K^\pm$ decay to three pions \cite{Batley:2006tt, Batley:2006mu, Batley:2007yfa}.
The experiment used simultaneous $K^+$ and $K^-$ beams with a momentum of
$60$~\gevp~ propagating along the same beam line. Data were collected in
2003-04, providing large samples of fully reconstructed $\kccc$ and $\kcnn$ 
decays.  Here we report the results from a study of a partial sample of 
$\sim 30.4 \times 10^6$ $\kcnn$ decays recorded in the second half of the
2004 run with the purpose of providing a new empirical, model-independent
parameterization of the $\kcnn$ Dalitz plot. This parameterization describes 
the $\kcnn$ experimental data with no distortions from instrumental effects,
such as resolution, geometrical acceptance and detection efficiency,
as they would be measured 
by a detector with full acceptance and ideal performance.
It could also be useful, therefore, in the
development of new theoretical formulations of rescattering effects in
$\kcnn$ decay, or in the refinement of existing ones. 

Rescattering effects are much smaller in $\kccc$ than in $\kcnn$ decay 
because the invariant mass of any two-pion pair is always  $\ge 2m_+$, hence
any cusp structure in $\kccc$ decay is outside the physical region. 
Indeed, a good fit to $4.709 \times 10^8$ $\kccc$ decays, 
also collected in this experiment, has been obtained
without the addition of rescattering terms \cite{Batley:2007md}. So, for
$\kccc$ decay the empirical parameterization of its Dalitz plot by a
series expansion \cite{Amsler:2008zz}, with the parameters given
in ref. \cite{Batley:2007md}, is still valid.

\section{Beam and detectors}
\label{beamdet}
The two simultaneous beams are produced by $400$~\geve~ protons impinging on
a 40 cm long Be target. Particles of opposite charge with
a central momentum of $60$~\gevp~ and a momentum band of $\pm 3.8\%$ produced
at zero angle are selected by a system of dipole magnets forming an
``achromat'' with null total deflection, focusing quadrupoles, muon sweepers
and collimators. With $7\times 10^{11}$ protons per burst of $\sim 4.5$ s 
duration incident on the target the positive (negative) 
beam flux at the entrance of the decay volume
is  $3.8\times 10^{7}$ ($2.6\times 10^{7}$) particles per pulse, of which
$\sim 5.7\%$ ($\sim 4.9\%$) are $K^+$ ($K^-$). The decay volume is a 114 m 
long vacuum tank with a diameter of 1.92 m for the first 66 m, and 2.4 m 
for the rest.

Charged particles from $K^\pm$ decays are measured by a magnetic spectrometer 
consisting of four drift chambers ($DCH$) and a large-aperture dipole magnet located
between the second and third chamber \cite{Fanti:2007vi}. Each chamber
has eight planes of sense wires, two horizontal, two vertical and two
along each of two orthogonal $45^\circ$ directions. The spectrometer
is located in a tank filled with helium at atmospheric pressure and
separated from the decay volume by a thin (0.0031 radiation lengths, $X_0$)
Kevlar window. A 16 cm diameter aluminium vacuum tube centered on the
beam axis runs the length of the 
spectrometer through central holes in the Kevlar window, drift chambers and 
calorimeters. Charged particles are magnetically deflected in the horizontal
plane by an angle corresponding to a transverse momentum kick of 
$120$ MeV/{\it c}. The momentum resolution of the spectrometer is 
$\sigma(p)/p = 1.02\% \oplus 0.044\%p$ ($p$ in~\gevp), as derived form the
known properties of the spectrometer and checked with the measured invariant
mass resolution of $\kccc$ decays. The magnetic spectrometer is followed by
a scintillator hodoscope consisting of two planes segmented into horizontal
and vertical strips and arranged in four quadrants.

A liquid Krypton calorimeter (LKr) \cite{Barr:1995kp} is used to reconstruct 
$\pi^0 \rightarrow \gamma \gamma$ decays. It is an almost homogeneous 
ionization chamber with an active volume of 
$\sim 10 ~m^3$ of liquid krypton, segmented transversally into 13248 
$2~ cm \times 2 ~cm$ projective cells by a system of
Cu-Be ribbon electrodes, and with no longitudinal segmentation. 
The calorimeter is 27 $X_0$ thick and has an energy resolution 
$\sigma(E)/E = 0.032/\sqrt{E} \oplus 0.09/E \oplus 0.0042$ (E in~\geve). 
The space resolution for a single electromagnetic shower can be parameterized as
$\sigma_x = \sigma_y = 0.42/\sqrt{E} \oplus 0.06$ cm for each transverse 
coordinate $x,y$.

A neutral hodoscope consisting of a plane of scintillating fibers is installed
in the LKr calorimeter at a depth of $\sim 9.5 ~X_0$. It is
divided into four quadrants, each consisting of eight bundles of vertical
fibers optically connected to photomultiplier tubes.

\section{Event selection}
\label{selection}

A specific subset (about 50\%) of the full data sample (collected in 2003 and 2004) 
was used, recorded with optimised trigger conditions allowing precise control 
of the trigger efficiency. 

$\kcnn$ events were recorded by a first level trigger using signals from the 
scintillator hodoscope (Q1) and LKr (NUT), followed by a second level trigger using drift
chamber information (MBX). Events were also recorded using other
triggers with different downscaling factors for different periods: a minimum bias
NUT trigger (ignoring both Q1 and MBX); and a minimum bias Q1*MBX trigger
(ignoring LKr information). Using the event samples recorded with these downscaled 
triggers, and selecting $\kcnn$ decays, it was possible to measure separately 
the efficiency of the minimum bias Q1*MBX trigger using the event sample 
recorded by the minimum bias NUT trigger and the efficiency of the minimum bias 
NUT trigger using the events recorded by the minimum bias Q1*MBX trigger. 
These two efficiencies were multiplied together to obtain the full trigger 
efficiency, which was always above 94\% for the data sample used in
this analysis. Details of the trigger 
efficiency for $\kcnn$  decay events are given in \cite{Batley:2006tt,Batley:2007yfa}.

Events with at least one charged particle track having a momentum above 
5~\gevp, measured  with a maximum error of 6\%,
and at least four energy clusters in the LKr, each consistent with a photon
and above an energy threshold of 3~\geve, were selected for further analysis. In addition, 
the distance between any two photons in the LKr was required to be larger
than 10~cm, and the distance between each photon and the impact point of any
track on the LKr front face had to exceed 15~cm. Fiducial cuts on the distance of each photon
from the LKr edges and centre were also applied in order to ensure 
full containment of the electromagnetic showers.

Every combination of four clusters and one track was considered as a $\kcnn$ decay 
candidate if clusters were in time within 5 ns, and if the track was in time with the cluster 
average time within 10 ns. The distribution of the difference between the 
time of each cluster and their average value has an approximately Gaussian 
shape with $\sigma \approx 0.73$ ~ns, while the distribution of the 
difference between the track time and the cluster average time has 
$\sigma \approx 1.5$~ns, so these cuts accept almost all the
time-correlated combinations. At this stage of event
selection there is a $\sim 1.5$\% background associated with
accidental LKr clusters. However, after the $\pi^0 \pi^0$ pair selection 
(see below) the level of residual accidental background, estimated
from the distribution of the difference between the 
track time and the average time of the four clusters, is less than
0.02\% and can be safely neglected. 

Other rate effects, such as losses caused by mismeasurement of cluster and track parameters due 
to accidental activity in the detectors, were considered as part of
the detector performance. The simulation of relevant resolutions and
tails has been tuned to the experimental data, hence our 
Monte Carlo model includes also these rate effects. Residual discrepancies
between experimental and simulated samples were taken into account in the
study of systematic uncertainties (see section \ref{systematics}).

\begin{sloppypar}
Each possible combination of two photon pairs in the event was assumed to originate from  
the two-photon decays of a pair of neutral pions, and for every $\pi^0$ candidate 
the position of the decay vertex along the beamline was calculated as
$Z_{\pi^0} = Z_{LKr} - \frac{\sqrt{E_1E_2((x_1 - x_2)^2+(y_1 - y_2)^2)}}{m_{\pi^0}}$,
where $Z_{LKr}$ is the LKr longitudinal position, and $E_1, E_2, x_1, x_2, y_1, y_2$ are the 
measured energies and transverse coordinates of the two photons, as
measured in the LKr.
The $\kcnn$ decay
vertex position $Z$ was taken as the arithmetic average of the two $Z_{\pi^0}$ values.    
The reconstructed decay vertex position $Z$ was further required to be
at least 2 m after the downstream end of the final beam collimator. In addition, the
reconstructed kaon momentum was required to be between 54 and 66~\gevp.
\end{sloppypar}

For each $DCH$ plane the event energy-weighted center-of-gravity (COG) 
coordinates were calculated using the photon coordinates and energies, as 
measured by the LKr, and the track parameters before deflection,
so COG represents the intersection of the initial kaon flight line
with the $DCH$ plane.
Inner acceptance cuts were applied at each $DCH$ plane to reject events with 
COG radius larger than $R^{COG}_{max}$ (typically between 2 and 3~cm)\footnote{The
beams were focused at DCH1, where the RMS values of their radial distributions
were $\sim 0.45 cm$}, 
and with the charged track closer than $R^{COG-track}_{min}$
(typically between 15.5 and 19~cm) to the event COG.
The exact cut values for every $DCH$ plane have been chosen depending  
on the COG and track impact point
distributions on that plane.

In order to reject events with photons emitted at very small angles to
the beam and traversing the beam pipe in the spectrometer or the DCH1
central flange, and converting to $e^+e^-$ before reaching the LKr, 
for each photon detected in LKr its distance from the nominal beam axis at the DCH1 
plane was required to be $> 11$ cm, assuming an origin on axis at $Z+400~cm$. 

\begin{sloppypar}
For every $\kcnn$ decay candidate in the event, both 
the reconstructed $\pi^{\pm} \pi^0 \pi^0$ invariant mass ($M$) and 
the difference between the two $Z_{\pi^0}$ coordinates ($\delta Z$)
were used. 
For each $\kcnn$ decay candidate an estimator $\chi^2$ was defined as 
$\chi^2 = (\delta Z/RMS_z(Z))^2 + ((M-M_{PDG})/RMS_m(Z))^2$, where the 
resolutions $RMS_z$ and $RMS_m$
have been parameterized from the experimental data as a functions of $Z$.
The combination with the minimum $\chi^2$ was chosen as
the reconstructed $\kcnn$ decay after applying  
the final loose cut $\chi^2 < 30$.
\end{sloppypar}

\begin{figure}[h]
\begin{center}

\resizebox{0.5\textwidth}{!}{%
\setlength{\unitlength}{1mm}
\begin{picture}(100.,100.)

\includegraphics[width=100mm]{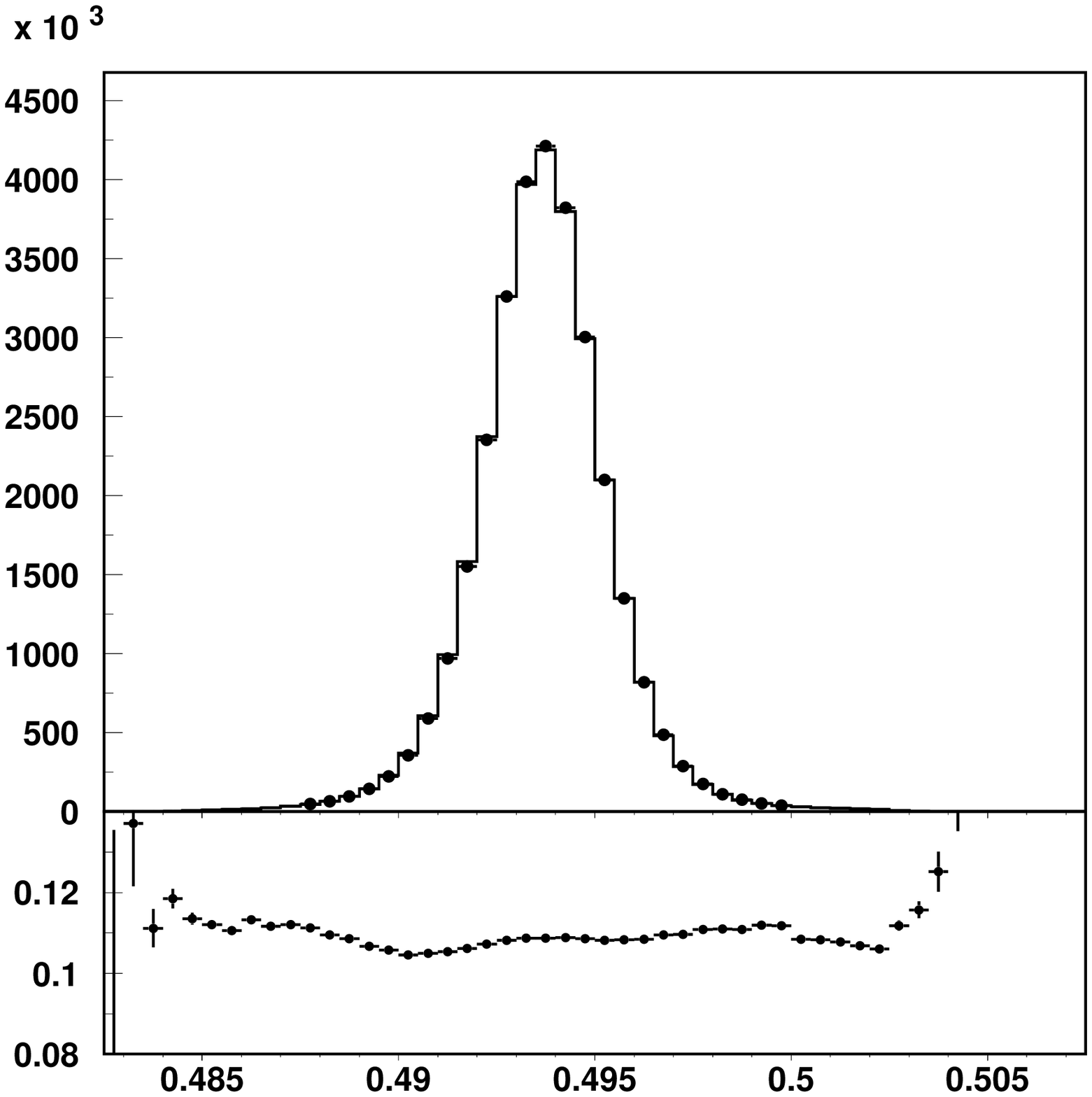}

\put(-50.,2.){\makebox(0,0){$M$ (\gevm)}}

\put(-14.,85.){\makebox(0,0){ a)}}
\put(-14.,25.){\makebox(0,0){ b)}}

\put(-103.,60.){\makebox(0,0){\rotatebox{90}{events/0.0005  \gevm}}}

\end{picture}
}

\end{center}

\caption{
Reconstructed $\pi^{\pm} \pi^0 \pi^0$ invariant mass ($M$) distributions for data and MC simulation. 
{\rm a)} Experimental (solid circles) and simulated (histogram) 
distributions, normalized to the number of data events. 
{\rm b)} Ratio between the experimental and simulated distributions
before this normalization.
}
\label{dmk}
\end{figure}

The $\pi^{\pm} \pi^0 \pi^0$ invariant mass distribution is shown in Fig. \ref{dmk}. 
Non-gaussian tails, mainly associated with $\pi \to \mu \nu$ decays
in $\kcnn$ events, are suppressed by the $\chi^2$ cut. There are also small 
contributions from wrong photon pairings in the decay of the two $\pi^0$, 
and from non-gaussian tails of the LKr response due to photonuclear reactions. 
All these effects are included in the Monte Carlo 
simulation and are taken into account in the evaluation of the
systematic uncertainties (Section \ref{systematics}).

Radiative photons from $\kcnn$ decays produce a slight shift of the measured
kaon mass, and thus also contribute to the tails of the $\chi^2$ distribution.
Our simulation does not take into account radiative photons, and we
assume that the emission of soft real $\gamma$ leaves the decay kinematics
essentially unchanged.
There is no limit to the presence of additional clusters in our event
selection from the data. We have checked 
that the replacement of the $\chi^2$ cut with the cut $\delta Z  <
500\ cm$ (with no cuts on the measured $\pi^{\pm} \pi^0 \pi^0$ invariant mass)
leads to a negligible change of the $s_3$ spectrum and of
the fit results. So, within the present statistical uncertainty our 
analysis includes all the radiative  $K^{\pm} \rightarrow \pi^\pm \gamma \pi^0 \pi^0$ decays.

There are no important physical background sources for the $\kcnn$ decay mode. 
Accidental overlaps of two events could produce some
background, which, however, is expected to have a flat distribution in
the $\delta Z, M$ plane, hence a flat $\chi^2$ distribution.
If one interprets the small differences observed in the tails of the $\chi^2$
distributions of data and MC events 
as totally due to this background rather than to the
quality of the simulation, the accidental background 
can be conservatively estimated to be $< 0.2\%$. 

\begin{figure}[h]
\begin{center}

\resizebox{0.5\textwidth}{!}{%
\setlength{\unitlength}{1mm}
\begin{picture}(100.,100.)
\epsfig{file=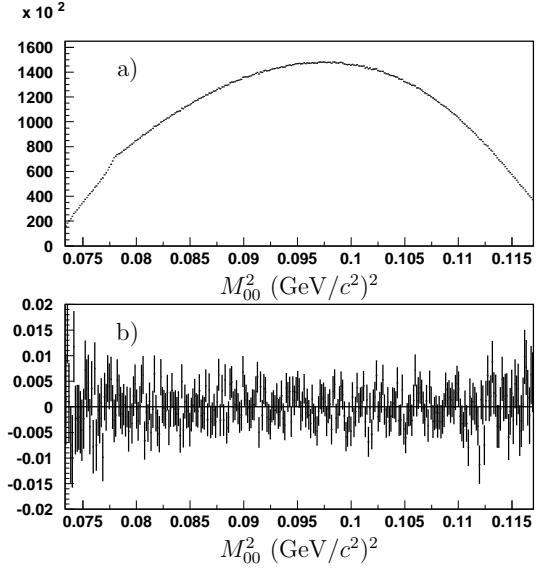,width=100mm}
\put(-80.,85.){\makebox(0,0){ a)}}
\put(-80.,40.){\makebox(0,0){ b)}}

\put(-50.,48.){\makebox(0,0){$\mm2$  \gevmsq}}
\put(-50., 3.){\makebox(0,0){$\mm2$  \gevmsq}}
\end{picture}
}

\end{center}
\vspace{-0.5cm}
\caption{
{\rm a)} - experimental distribution of the square of the $\pi^0\pi^0$
mass, $\mm2$, from $\kcnn$ decay in the fit region.
{\rm b)} - relative deviation of the experimental spectrum from the best fit result
$(Data-Fit)/Fit$. 
}
\label{deviation}
\end{figure}

A total of $30.4 \times 10^6$ $\kcnn$ decay candidates have been selected for the present 
analysis. Fig. \ref{deviation} a) shows the distribution of the square of the $\p0p0$ 
invariant mass, $\mm2$, for the final event sample. This distribution is 
displayed with a bin width of 0.00015~\gevmsq, with the $51^{st}$
bin centered at $\mm2 = (2m_+)^2$, where $m_+$ is the charged pion mass 
(the $\mm2$ resolution is 0.00031~\gevmsq~at $\mm2 = (2m_+)^2$). 
For our fits we use the bin interval $21-311$ 
which contains the major part ($> 98\%$) of selected events. The sudden change of 
slope near $\mm2 = (2m_+)^2 = $ 0.07792~\gevmsq, first observed 
in this experiment \cite{Batley:2005ax} is clearly visible.

\section{Monte Carlo simulation}
\label{mcsimul}
Samples of simulated $\kcnn$ events $\sim10$ times larger than the data have
been generated using a full detector simulation based on the GEANT-3 package
\cite{Brun:1978fy}. This Monte Carlo (MC) program takes into account
all known detector effects, including the time-dependent
efficiencies and resolutions of the detector components. 

The MC program also includes the simulation of the beam line.
The beam average position and momentum are tuned for each period of few hours using 
fully reconstructed $\kccc$ events, 
which provide precise information on the average beam angles and positions. Furthermore, 
the requirement that the average reconstructed $\pi^\pm \pi^+ \pi^-$ invariant mass be equal 
to the nominal $K^\pm$ mass for both $K^+$ and $K^-$ fixes the absolute momentum scale 
of the magnetic spectrometer.

The Monte Carlo simulation does not include the overlay of two
independent $\kcnn$ events or of a simulated $\kcnn$ event with a randomly triggered one, 
so the timing cuts described in section \ref{selection} were not
applied in the analysis of the simulated event sample. 
It should be noted that rate effects depend on
the time structure of the SPS beam spills, which may vary from spill
to spill during data taking and cannot be easily included in the Monte Carlo simulation.

The Dalitz plot distribution of $\kcnn$ decays has been generated according to Eq.(\ref{old}).
For any given value of the generated $\p0p0$ invariant mass the simulation provides 
the detection probability and the distribution function for the reconstructed value of $\mm2$. 
This allows the transformation of any theoretical distribution into an expected 
distribution which can be compared directly with the measured one.

\section{Parameterization}
\label{paramsec}

In order to describe the cusp observed in the $\pi^0\pi^0$ invariant mass distribution, 
we propose the following empirical parameterization for the square of
the $\kcnn$ decay matrix element:
\begin{equation} 
\frac{d |M|^2 }{dU dV} \propto [M_u(U) +  \frac{k V^2}{2}]^2 f(U),
\label{parameq}
\end{equation}
where 
\begin{eqnarray}
\nonumber
M_u(U) = 1 + \frac{g U}{2}  + \frac{h U^2}{2} + \\
 a (U_t-U)^q H (U_t-U) + b (U-U_t)^q H(U-U_t),
\label{udep}
\end{eqnarray} 
and
\begin{equation}
f(U) = 1 + p w \delta (U - U_t).
\label{upeak}
\end{equation}

\begin{sloppypar}
Here $H$ is the Heaviside step function (\mbox{$H(x < 0)=0$}, \mbox{$H(x \ge 0)=1$}) 
and $\delta$ is the Dirac delta function (in particular, \mbox{$\int_{-w/2}^{+w/2} \delta(x)dx=1$}). 
The constant $U_t= -1.1272$ is the $U$ value at the threshold of charged pion pair 
production, which corresponds to $s_3 = 4 m_{\pi^+}^2$. The factor $f(U)$ takes into account     
the additional contribution from $\pi^+ \pi^-$ bound states and other narrow peaks from
electromagnetic effects, all decaying to $\pi^0 \pi^0$ \cite{Gevorkyan:2006rh}. 
All these contributions have widths that are much narrower than our experimental $\mm2$ 
mass resolution.
\end{sloppypar}

The $s_3$ bin width used to store the measured spectrum is denoted as $w$.   
With this definition the parameter $p$ is dimensionless, and represents the relative increase
of the content of the bin containing the value $U = U_t$ with respect to the value
calculated with $f(U) = 1$.  In our
analysis we use $w$ = 0.00015~\gevmsq~, and all the $p$ values listed
below are written for this value of bin width.

The exponent $q$ could be different, in principle, above and below the cusp point,
but our fits show that there is no need for such an additional degree of freedom,
because the $\mm2$ shape in these two regions is successfully described by the two 
independent constants $a$ and $b$.

The parameters describing the $\kcnn$ Dalitz plot are
$g,h,k,a,b,p,q$. The parameters $g,h,k$ are not equivalent to the
corresponding constants $G,H,K$ of the old PDG parameterization (\ref{old}) \cite{Amsler:2008zz}
and of the physical parameterizations \cite{Batley:2009cusp}, but have
a similar meaning. The expression (\ref{udep}) is inspired by the Cabibbo-Isidori physical parameterization of the 
$\kcnn$ matrix element at tree level \cite{Cabibbo:2004gq,Cabibbo:2005ez}. 
The last two terms of (\ref{udep}) correspond to an empirical description  
of the $\pi\pi$ rescattering effects \cite{Cabibbo:2005ez,Colangelo:2006va,Bissegger:2008ff}.

\section{Fitting the data}

The $V$-dependence of expression (\ref{parameq}) is described by the $\frac{k V^2}{2}$ term which
is known from earlier measurements to be rather small, $k \approx 0.01$ 
\cite{Ajinenko:2002mg, Akopdzhanov:2005nb, Batley:2009cusp}. So, ignoring the term $\propto k^2$, 
the $U$-dependence of the $\kcnn$ decay width can be expressed as
\begin{eqnarray}
\nonumber
\frac{d \Gamma}{ d U} \propto \int_0^{V_{max}(U)} \frac{d |M|^2 }{dU dV}  dV = \\
= V_{max}(U) f(U)(M_u^2 + \frac{1}{3}M_u k V_{max}^2(U)),
\label{vterm}
\end{eqnarray}
where $V_{max}(U)$ is the maximum kinematically allowed $V$ for a given $U$. If $k$ is known,
formula (\ref{vterm}) can be used to fit the $\kcnn$ decays $U$-distribution provided the
sensitivity of the acceptance to the small $\frac{k V^2}{2}$ term 
is taken into account as a contribution to the systematic uncertainty of the results.

\begin{sloppypar}
The main $U$-dependence of the parameterization is described by the parameters 
$g,h,a,b,p,q$, 
which are related to the measurement of  $s_3$, which is equal to the the square of the 
$\pi^0\pi^0$ invariant mass, $\mm2$. So the systematic uncertainties of these parameters 
depend mainly on the performance of the LKr calorimeter. The measurement of $k$ relies also on the 
measurement of the $\pi^\pm$ track in the DCH, but due to the smallness of the $k$ value 
its uncertainty affects only weakly the determination of the other parameters. 
Furthermore, $a,b,p,q$ describe the fine features 
of the Dalitz plot in the cusp region that require narrow $s_3$ bins,
while the $k$ term of formula (\ref{vterm}) is smooth over the Dalitz plot and does not require
such narrow bins.  So we have decided to measure the
$V$-dependence of the Dalitz plot separately by an iterative procedure.
\end{sloppypar}

Assuming an initial value $k = 0.01$, a first fit to the one-dimensional $s_3$ distribution has been 
performed using the MINUIT package. The $\chi^2$ was calculated from
the difference between the number of observed events in each bin and the
number predicted from the parameterization (\ref{vterm}) with the current values of the fit parameters.
The predicted number of events was calculated by convoluting the parameterization 
(\ref{vterm}) 
with the MC distributions of the measured $s_3$ for each generated ('true') $s_3$ value. 
In such a way both acceptance and resolution effects were taken into account. 

The parameters $a,b,p,q$, describing the cusp shape, were then fixed
to the values obtained from the first fit and used for the two-dimensional fit to 
determine $k$. This fit was performed by implementing the event-weighting technique in the MINUIT package. 
As a first step, the number of events in each bin of the experimental Dalitz plot was corrected 
for the trigger inefficiency. Then, at each step of the $\chi^2$ minimization
the full MC sample corresponding to the experimental data used in the fit 
($\approx 280 \times 10^6$ events) was used to build a simulated
Dalitz plot by giving each event a weight equal to the ratio between the parameterization (\ref{parameq}) 
with the current values of the fit parameters, and (\ref{old}), which was used for the simulation of
MC events. In the calculation of the weights the 'true' $U,V$ values
were used, while the MC events were binned using the reconstructed $U,V$ values (here $V$ means $|V|$).
The MC Dalitz plot was normalized to the total number of data events.
The $\chi^2$ was then calculated from the difference between the MC and data Dalitz plots.

For the two-dimensional $U,V$ histograms we used $50 \times 50$ bins in the intervals
$-1.45<U<1.35$ , $0<V<2.8$. The $\chi^2$ contribution
was calculated for the center of each bin over the $U$  
range corresponding to the one-dimensional fit limits, and with the $V$ upper limit set to 
$0.9 V_{max}(U)$ to avoid tails effect.

\begin{figure}[h]
\begin{center}

\resizebox{0.5\textwidth}{!}{%
\setlength{\unitlength}{1mm}
\begin{picture}(100.,100.)
\epsfig{file=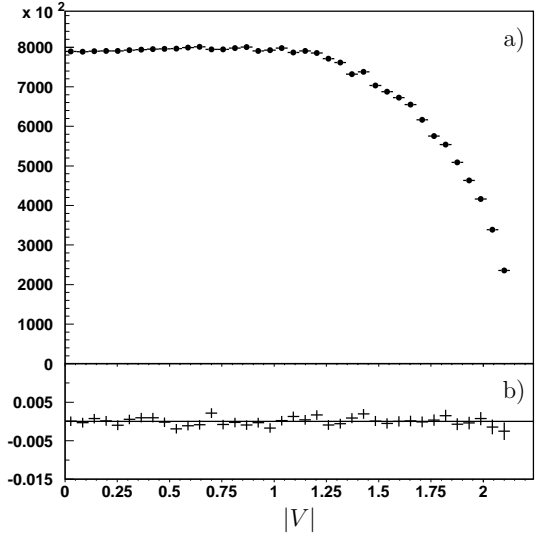,width=100mm}
\put(-14.,85.){\makebox(0,0){ a)}}
\put(-14.,25.){\makebox(0,0){ b)}}

\put(-50., 3.){\makebox(0,0){$|V|$}}
\end{picture}
}

\end{center}
\vspace{-0.5cm}
\caption{
{\rm a)} - experimental $|V|$ distribution obtained by projection of the ($U,|V|$)
distribution used in the two-dimensional fit to extract the $k$ parameter, after 
correction for the trigger inefficiency. 
{\rm b)} - Deviation from 1 of the ratio between the experimental and 
normalized simulated distributions for the best two-dimensional fit parameters.
}
\label{vcompare}
\end{figure}

This fit was performed with $a,b,p,q$ fixed to the values obtained from the
one-dimensional fit made with formula (\ref{vterm}) under 
the initial assumption $k=0.01$. The result of the two-dimensional fit was $k = 0.0081(2)$. When
the procedure was repeated with $k=0.0081$ as the initial assumption, it reproduced the measurement
$k=0.0081(2)$ with $\chi^2 = 1163.5$ for 1249 degrees of freedom (probability 0.96), 
so no further iteration was needed. 
The fit without the trigger correction gives $k=0.0086(2)$, providing an estimate of the trigger inefficiency effect, 
which is conservatively taken as the contribution to the systematic error on $k$. Thus, our
result for the $k$ parameter of the Dalitz plot is
\begin{equation}
k = 0.0081 \pm 0.0005_{Trigger} \pm 0.0002_{Stat} = 0.0081(5).
\label{kres}
\end{equation}
Fig. \ref{vcompare} shows the comparison of the experimental and
simulated $|V|$ distributions obtained by projection of the 
the two-dimensional distribution used in the fit to extract the $k$ value (\ref{kres}).

Using the fixed $k$ value given in (\ref{kres}), the values of all other parameters in (\ref{parameq})
as well as their systematic uncertainties were obtained from the fit to the one-dimensional $s_3$ 
distribution, after correcting the content of each bin for the trigger efficiency.
The fit gives $\chi^2=265.1$ for 284 degrees of freedom (probability 0.78). 
The best fit values of the parameters are listed in Table \ref{empiric}. 
The uncertainty affecting the $k$ value is taken into account as one of the sources
of systematics errors for the other parameters, and is denoted as $k$
error in Table \ref{empiric}. The effect of the trigger efficiency is also conservatively 
taken as the contribution to the systematic error for every parameter,
and is denoted as Trigger in Table \ref{empiric}.

\begin{table*}[ht]
\caption{Contributions to systematic uncertainties, statistical errors and central values for 
the empirical fit parameters}
\begin{center}
\begin{tabular}{|l|l|l|l|l|l|l|} \hline
                      &    $g$   &    $h$   &    $a$   &    $b$   &    $p$   &    $q$   \\ \hline  
Acceptance(Z)         &   0.0052 &   0.0043 &   0.0021 &   0.0029 &   0.0096 &   0.0177 \\ 
Acceptance(V)         &   0.0002 &   0.0002 &   0.0004 &   0.0005 &   0.0018 &   0.0033 \\ 
LKr resolution        &   0.0009 &   0.0012 &   0.0002 &   0.0009 &   0.0068 &   0.0009 \\ 
LKr non-linearity     &   0.0089 &   0.0086 &   0.0038 &   0.0075 &   0.0250 &   0.0406 \\ 
$P_K$ spectrum        &   0.0000 &   0.0003 &   0.0001 &   0.0004 &   0.0006 &   0.0001 \\ 
MC(T)                 &   0.0005 &   0.0006 &   0.0003 &   0.0006 &   0.0040 &   0.0034 \\ 
Trigger               &   0.0027 &   0.0052 &   0.0051 &   0.0037 &   0.0065 &   0.0260 \\
$k$ error             &   0.0004 &   0.0004 &   0.0002 &   0.0002 &   0.0006 &   0.0014 \\ 
Hadronic showers      &   0.0005 &   0.0004 &   0.0003 &   0.0003 &   0.0017 &   0.0028 \\ \hline
Systematic error      &   0.0107 &   0.0110 &   0.0067 &   0.0089 &   0.0288 &   0.0517 \\ \hline 
Statistical error     &   0.0013 &   0.0014 &   0.0031 &   0.0026 &   0.0145 &   0.0204 \\ \hline
Total uncertainty     &   0.0108 &   0.0111 &   0.0074 &   0.0093 &   0.0322 &   0.0556 \\ \hline 
Central value        &   0.6715 &  -0.0270 &  -0.1299 &  -0.0378 &   0.0661 &   0.4474 \\ \hline
\end{tabular}
\end{center}
\label{empiric}
\end{table*}

\section{Systematic uncertainties}
\label{systematics}

All sources of systematic uncertainties are described in detail in ref. \cite{Batley:2009cusp}.

The detector acceptance to $\kcnn$ decays depends strongly on the 
position of  the $K^{\pm}$ decay vertex along the nominal beam axis, $Z$.
A small difference between the shapes of the experimental and simulated
distributions is present in the high $Z$ region (close to the spectrometer)
where the acceptance drops 
because of the increasing probability for the charged pion track 
to cross the spectrometer too close to the event COG. The effect of this
difference has been checked by introducing a small mismatch in the track 
radius cuts between real and simulated samples, and also 
by applying a small change to the  LKr energy scale (that leads to 
a shift of the measured $Z$ position). The corresponding small changes of the fit results 
are considered as the acceptance related contribution to the systematic errors 
(denoted as Acceptance(Z) in Table \ref{empiric}).

The simulated sample from which the acceptance and resolution effects used
in the fits are derived, is generated under the assumption that the $\kcnn$ 
matrix element does not depend on $V$. We have studied the sensitivity
of the fit results to the presence of the $V$-dependent term compatible with our data 
in the simulated sample. The largest variations of the fit results are shown in 
Table \ref{empiric} as the contributions to the systematic errors arising from the 
simplified matrix element used in the MC (they are denoted as
Acceptance(V)). 

The $\pi^0 \pi^0$ invariant
mass, $M_{00}$, is determined using only information from the LKr
calorimeter. We find that a convenient variable which is sensitive to all random fluctuations 
of the LKr response, and hence to its energy resolution, is the ratio 
$m_{\pi^0_1}/m_{\pi^0_2}$, 
where $m_{\pi^0_1}$ and $m_{\pi^0_2}$ are the measured two-photon
invariant masses for the more and less energetic $\pi^0$, respectively,
in the the same event. The width of
the distribution for simulated events is slightly larger than that of
the data: the RMS value of the simulated distribution is 0.0216, while it is 0.0211 
for the data. 

In order to check the sensitivity of the fit results to  
a resolution mismatch of this size, we have smeared the measured
photon energies in the data by adding a random energy with a gaussian distribution
centered at zero and with $\sigma = 0.06$~\geve. Such
a change increases the RMS value of the $m_{\pi^0_1}/m_{\pi^0_2}$
distribution from 0.0211 to 0.0224. A fit is then performed for the data 
sample so modified, and the values of the fit parameters are compared with
those obtained using no energy smearing. 

The artificial smearing of the photon energies described above introduces random  
shifts of the fit parameters within their statistical errors. In order
to determine these shifts more precisely than allowed by the
statistics of a single fits, we have repeated the fit eleven times
using for each fit a data sample obtained by smearing the original one
with a different series of random numbers. The shifts of the fit parameters, 
averaged over the eleven fits, are then taken to represent the
systematic effects, while the errors on the average values are the corresponding
uncertainties. The shifts and their errors so defined, summed quadratically, 
are denoted as ``LKr resolution'' in the list of systematic errors
given in Table \ref{empiric}.   

In order to study possible non-linearity effects of the LKr calorimeter response
to low energy photons, we select $\pi^0$ pairs from $\kcnn$ events with
symmetric $\pi^0 \rightarrow \gamma \gamma$ decays  ($0.45 < E_{\gamma}/E_{\pi^0} < 0.55$),
and with the more energetic $\pi^0$ (denotes as $\pi^0_1$) in the energy range 
$22\ \mathrm{GeV} < E_{\pi^0_1} < 26\ \mathrm{GeV}$.

For the $\pi^0$ pairs so selected we define the ratio of the
two-photon invariant masses, $r=M_{\pi^0_2}/M_{\pi^0_1}$, where $\pi^0_2$
is the lower energy $\pi^0$.
Because of the resolution effects discussed above its average value 
$\langle r\rangle$ depends on the lower pion energy even in the case of perfect LKr linearity.
However, for $E_{\pi^0_2}/2 < 9$~GeV  
the values of $\langle r\rangle$ for simulated events are systematically above
those of the data, providing evidence for the presence of non-linearity   
effects of the LKr response at low energies.

To study the importance of these effects, we modify all simulated events to 
account for the observed non-linearity multiplying each photon
energy by the ratio $\langle r_{Data}\rangle/\langle r_{MC}\rangle$, where $\langle r_{Data}\rangle$
and $\langle r_{MC}\rangle$ are the average ratios for data and simulated events, respectively. 
The values of $\langle r\rangle$ for the sample of simulated
events so modified are very close to those of the data. The small shifts  
of the best fit parameters obtained using these non-linearity corrections 
are taken as contributions to the systematic errors in Table \ref{empiric}, 
where they are denoted as ``LKr non-linearity''. 

The $\pi^{\pm}$ interaction in LKr may produce multiple energy
clusters which are located, in general, near the impact point of the $\pi^{\pm}$
track and in some cases may be identified as photons. To reject such
``fake'' photons a special cut on the distance $d$ between each photon and
the impact point of any charged particle track at the LKr 
is implemented in the event selection.
In order to study the effect of these ``fake'' photons on the best fit
parameters we have repeated the fits by varying the cut on the
distance $d$ between 10 and 25 cm in the selection of both data and
simulated $\kcnn$ events. The largest deviations from the results
obtained with the default cut value ($d$=15 cm) are taken as contributions 
to the systematic errors (see Table \ref{empiric}, ``Hadronic showers'').

The MC program includes a complete simulation of the beam magnet
system and collimators with the purpose of predicting the correlation between
the incident $K^{\pm}$ momenta and trajectories. However, the absolute beam
momentum scale cannot be predicted with the required precision, hence
we tune the average value to the measured ones for each continuous
data taking period (``run'') using $\kccc$ events which are recorded    
continuously during data taking, and also simulated by the MC
program.

After this adjustement, a residual difference still exists between the
measured and simulated $K^{\pm}$ momentum distributions. 
In order to study the sensitivity of the best fit parameters to this distribution, 
we have corrected the simulated momentum distribution to reproduce the 
measured one. The corresponding changes of the best fit parameters are 
included in the contributions to the systematic errors and denoted as 
'$P_K$ spectrum' in Table \ref{empiric}.  

In order to take into account variations of running conditions during
data taking, the number of simulated $\kcnn$ events for each run should be
proportional to the corresponding number of events in the data. However,
because of small variations of trigger efficiency and acceptance, 
the ratio between the number of 
simulated and real events varies by a few percent during the whole
data taking period. In order to study the effect of the small mismatch
between the two samples on the best fit parameters, we have made them
equal run by run by a random rejection of selected events. The corresponding
shifts of the best fit parameters are considered as a MC     
time dependent systematic error, and are listed in Table \ref{empiric},
where they are denoted as ``MC(T)''.

Correlations between the fit parameters are changed by the systematic
uncertainties from the values shown in the Table \ref{matrix0} (purely
statistical correlations) to the ones of Table \ref{matrix}.

 \begin{table}[!h]
 \caption{Correlation matrix for the statistical errors}
\begin{center}
 \resizebox{0.48\textwidth}{!}{%
 \begin{tabular}{|l|l|l|l|l|l|l|} \hline
   &    $g$  &    $h$  &    $a$  &    $b$  &    $p$  &    $q$  \\ \hline
 $g$ &   1.000 &             &             &             &             &             \\
 $h$ &   0.440 &   1.000 &             &              &             &             \\
 $a$ &  -0.886 &  -0.502 &   1.000 &             &             &             \\
 $b$ &   0.327 &   0.861 &  -0.434 &   1.000 &             &             \\
 $p$ &   0.297 &   0.518 &  -0.329 &   0.702 &   1.000 &             \\
 $q$ &   0.883 &   0.544 &  -0.915 &   0.619 &   0.508 &   1.000 \\
 \hline \end{tabular}
 }
\end{center}
 \label{matrix0}
 \end{table}

 \begin{table}[!h]
 \caption{Correlation matrix for the total uncertainties}
\begin{center}
 \resizebox{0.48\textwidth}{!}{%
 \begin{tabular}{|l|l|l|l|l|l|l|} \hline
   &    $g$  &    $h$  &    $a$  &    $b$  &    $p$  &    $q$  \\ \hline
 $g$ &   1.000 &             &             &             &             &             \\
 $h$ &   0.850 &   1.000 &             &             &             &              \\
 $a$ &  -0.839 &  -0.686 &   1.000 &             &             &             \\
 $b$ &   0.895 &   0.872 &  -0.728 &   1.000 &             &             \\
 $p$ &   0.820 &   0.758 &  -0.675 &   0.921 &   1.000 &             \\
 $q$ &   0.931 &   0.796 &  -0.903 &   0.917 &   0.855 &   1.000 \\
 \hline \end{tabular}
 }
\end{center}
 \label{matrix}
 \end{table}

\section{Conclusion}
\label{conclusion}

The square of the $\kcnn$ matrix element can be written using the empirical approximation
\begin{eqnarray}
\nonumber
\frac{d |M|^2 }{dU dV} \propto 
[1 + \frac{g U}{2}  + \frac{h U^2}{2} + \frac{k V^2}{2} + \\
\nonumber
+ a(U_t-U)^q H(U_t-U) + b(U-U_t)^q H(U-U_t)]^2 \cdot \\
 (1 + p w \delta (U - U_t)),
\end{eqnarray}
where $w$ = 0.00015~\gevmsq~ and 
$U_t = (4 m_{\pi^+}^2 - s_0)/m_{\pi^+}^2$  
with the following parameter values: 

$$g =  0.672   \pm 0.001_{Stat}    \pm 0.011_{Syst}    =   0.672 \pm 0.011$$ 
$$h = -0.027   \pm 0.001_{Stat}    \pm 0.011_{Syst}    =  -0.027 \pm 0.011$$
$$k =  0.0081 \pm  0.0002_{Stat.} \pm 0.0005_{Syst}  =   0.0081 \pm 0.0005$$
$$a = -0.130   \pm 0.003_{Stat}     \pm 0.007_{Syst}    =  -0.130 \pm 0.007$$
$$b = -0.038   \pm 0.003_{Stat}     \pm 0.009_{Syst}    =  -0.038 \pm 0.009$$
$$p =  0.07     \pm 0.01_{Stat}       \pm 0.03_{Syst}     =   0.07 \pm 0.03$$
$$q =  0.45     \pm 0.02_{Stat}       \pm 0.05_{Syst}     =   0.45 \pm 0.06$$

Near the cusp point $U = U_t$ this approximation is only valid if the $s_3$
distribution is averaged over bins which are wider than the intrinsic width
of the peak expected from $\pi^+ \pi^-$ bound states and other
electromagnetic effects \cite{Gevorkyan:2006rh}, all decaying to $\pi^0 \pi^0$.
This peak is much narrower than the bin width used here, $w$ = 0.00015~\gevmsq,  
which is of the order of the experimental resolution. 

The errors are dominated by systematic effects. The systematic errors
on the slope parameters $g,h$ are substantially larger than the errors on $g,h$ 
obtained from our study of the $\pi\pi$ scattering lengths based on 
the full 2003-2004 data sample \cite{Batley:2009cusp}. This is mainly because
we use here the almost full fit interval in order to give a complete description 
of the $\kcnn$ Dalitz plot, while 
the fitting range used in ref. \cite{Batley:2009cusp} was optimized to reach the smallest
total error for the measured $\pi\pi$ scattering lengths.
The wide $s_3$ fitting range increases the sensitivity of the results to LKr 
non-linearity and trigger inefficiency.

Finally, we note that there is no model-independent relation between the values of the
best fit parameters given above and the S-wave $\pi\pi$ scattering
lengths $a_0$ and $a_2$, which are meaningful variables only within a
specific formulation of $\pi \pi$ rescattering effects in $\kcnn$
decay (see \cite{Batley:2009cusp}). The empirical parameterization proposed 
here provides a good description of this decay mode, but makes no assumption 
about the physics mechanisms responsible for the observed cusp structure.

\section{Acknowledgements}
\label{acknow}
We gratefully acknowledge the CERN SPS accelerator and beam-line staff
for the excellent performance of the beam. We thank the technical staff of
the participating laboratories and universities for their effort in the
maintenance and operation of the detectors, and in data processing.


\section*{References}
\bibliographystyle{epj}
\bibliography{na48ecusp}

\end{document}